\documentclass{aastex}
\usepackage{spr-astr-addons}
\usepackage{url}

\DeclareGraphicsExtensions{.eps,.bmp,.wmf,.jpg,.pdf}

\RequirePackage{color}

\newcommand{\emaila}{alexanderoliveros@mail.uniatlantico.edu.co, marioacero@mail.uniatlantico.edu.co}

\begin{document}

\title{New holographic dark energy model with non-linear interaction}
\shorttitle{New holographic dark energy model}
\shortauthors{A.~Oliveros, M.~A.~Acero}

\author{A. Oliveros} \and \author{Mario A. Acero}
\affil{Programa de F\'isica, Universidad del Atl\'antico, Km 7 antigua v\'ia Puerto Colombia, Barranquilla, Colombia}
\email{\emaila}

\begin{abstract}
In this paper the cosmological evolution of a holographic dark energy model with a non-linear interaction between the dark energy and dark matter components in a FRW type flat universe is analysed. In this context, the deceleration parameter $q$ and  the equation state $w_{\Lambda}$ are obtained. We found that, as the square of the speed of sound remains positive, the model is stable under perturbations since early times; it also shows that the evolution of the matter and dark energy densities are of the same order for a long period of time, avoiding the so--called coincidence problem. We have also made the correspondence of the model with the dark energy densities and pressures for the quintessence and tachyon fields. From this correspondence we have reconstructed the potential of scalar fields and their dynamics.
\end{abstract}

\keywords{dark energy; cosmology}


\section{Introduction}
One of the biggest unsolved mysteries in physics is that of dark energy. Dark energy was proposed by several research groups back in 1998 aiming to explain the accelerated expansion regime governing the current universe. Later, this accelerated expansion was discovered thanks to the vast amount of astrophysical information obtained from measurements of the supernovae type Ia (SNIa) luminosities [\cite{riess,Perlmutter1997zf,Perlmutter1998np,Astier2005qq}], as well as to the anisotropies observed in the cosmic microwave background (CMB) radiation [\cite{Spergel2003cb,Spergel2006hy}], and the analysis of the large scale structure (LSS) of the universe [\cite{Tegmark2003ud,Abazajian2004aja}]. 

The cosmological constant is the most common and simple candidate to serve as dark energy [\cite{Weinberg1988cp,Padmanabhan2002ji,Sahni1999gb}], but this proposition contains several difficulties as the fine tuning and the coincidence problems. A number of models have been proposed in the literature including scalar field as an alternative to explain dark energy nature, among which one finds quintessence, k-essence, tachyon and dilaton fields [\cite{Copeland2006wr}]. Other approaches include brane-world models [\cite{Deffayet2001pu,Sahni1999gb}], modified gravity theories ($F(R), F(G), F(R,G)$ gravitation), where dark energy emerges as a modification of geometry [\cite{Carroll2003wy,Capozziello2003tk,Nojiri2006ri,Nojiri2008nt}] (for an exhaustive discussion on these scenarios, the reader may review Ref. [\cite{Nojiri2010wj}]), and dark energy models with unconventional equation of state [\cite{Kamenshchik2001cp,Nojiri2005sr}]. 

Recently, a different idea has attracted considerable attention, proposing a model which has its roots on quantum gravity and is known as the holographic principle [\cite{Bekenstein1973ur,Bekenstein1974ax,Bekenstein1980jp,Bekenstein1993dz,Hawking1974sw,Hawking1976de,tHooft1993gx,Bousso1999xy,Cohen1998zx,Susskind1994vu}]. The foundation of the holographic principle resides in the idea that the entropy of a given system does not depend upon the volume, but rather on the surface area surrounding it. In a cosmological context, the holographic principle establishes an upper limit for the entropy of the universe [\cite{Fischler1998st}]. 

The authors of \cite{Cohen1998zx}, conjecture that in a quantum field theory there exists a relationship between an ultraviolet (UV) cutoff and an infrared (IR) cutoff, $L$, which origins lies on the constraint imposed to a region of space preventing a black hole formation. The size of this region should not be larger than $L$, meaning that if there is an energy density $\rho_{\Lambda}$ (zero-point quantum energy density) in a region associated to the UV, then the total energy in a region of size $L$ cannot overpass the mass of a black hole with the same size, so that
\begin{equation}\label{ecu1}
L^3\rho_{\Lambda}\leq LM_p^2.
\end{equation}
In the cosmological context, the holographic energy density $\rho_{\Lambda}$ corresponds to the dark  energy density (holographic dark energy) [\cite{Cohen1998zx,Hsu2004ri,Li2004rb}]. The maximum value $L$ can take is obtained by considering the equality in equation (\ref{ecu1}), such that the holographic energy density is given by
\begin{equation}\label{ecu2}
\rho_{\Lambda}=3c^2M_p^2L^{-2},
\end{equation}
where $c$ is a numerical constant and $M_p^{-2}=8\pi G$ is the reduced Planck mass.

Several approaches for the IR cutoff $L$ have been proposed in the literature [\cite{Li2004rb,Gao2007ep,Granda2008dk}]. For instance, in [\cite{Nojiri2005pu}], a generalized holographic dark energy model was proposed, identifying the IR cutoff with combination of FRW parameters. In this work, we consider the one given by Granda and Oliveros in \cite{Granda2008dk}, where the authors took the holographic dark energy density to depend on the usual quadratic term in the Hubble parameter ($H$) as well as on the time derivative of $H$, i.e., $\rho_\Lambda=3M_p^2(\alpha H^2+\beta \dot{H})$. This model has been widely studied by the scientific community in various contexts and it has been known as the new holographic dark energy model (NHDE).

Among the vast variety of approaches to the dark energy problem, it is also considered the possibility to include an interaction term $Q$ between the dark energy and dark matter components, usually introduced aiming to avoid the coincidence problem (in \cite{Bolotin2013jpa} a recent review of this topic is offered). Recently, the authors of \cite{Salvatelli2014zta} showed that this kind of interaction is favoured by current cosmological observations in an extended $\Lambda$CDM model. Due to our ignorance about the fundamental origin of dark energy, the interaction term is introduced by phenomenological considerations. This term is usually proportional either to the dark energy density, the dark matter or to a linear combination of both (see for instance, \cite{Arevalo2011hh} and references therein). The sign in $Q$ determines the direction of the flux of energy. If it is $Q$ is positive, dark energy decays in dark matter, while if $Q$ is negative, dark matter decays on dark energy. Some authors have also considered the possibility to have non-linear interaction terms [\cite{Arevalo2011hh,Mangano2002gg,Baldi2010vv,Ma2009uw,Zhang2012uu}]. Here we take this last idea, whit a concrete selection for the interaction term $Q$, which allows us to find analytical and phenomenologically feasible solutions.

This article is organised as follows: in section 2 we introduce the dark energy holographic model together with the interaction term between dark matter and dark energy, and we obtain the basic quantities to be used in subsequent calculations. Section 3 shows the correspondence of this model with quintessence and tachyon scalar fields. Finally, in section 4 we expose our conclusions.

\section{The Model}
The Friedmann equations for a flat Universe type FRW are
\begin{equation}\label{eq1}
H^2=\frac{1}{3}(\rho_m+\rho_{\Lambda}),
\end{equation}
\begin{equation}\label{eq2}
\dot{H}+H^2=-\frac{1}{6}(\rho_m+\rho_{\Lambda}+3p_{\Lambda}),
\end{equation}
where we take $8\pi G = 1$ and $p_m = 0$; $\rho_m$ is the dark matter density, $\rho_{\Lambda}$ is the dark energy density and $p_{\Lambda}$ the pressure due to the dark energy component. $H = \dot{a}/a$ is the Hubble parameter and the dot represents the derivative with respect to time. We considered that the dark energy component is the holographic energy density which has the form [\cite{Granda2008dk}]
\begin{equation}\label{eq3}
\rho_{\Lambda}=3(\alpha H^2+\beta \dot{H}),
\end{equation}
where $\alpha$ and $\beta$ are positive constants.

When a possible interaction between dark matter and dark energy components are taken into account, the continuity equations take the form
\begin{equation}\label{eq4}
\dot{\rho}_{m}+3H\rho_m=Q,
\end{equation}
\begin{equation}\label{eq5}
\dot{\rho}_{\Lambda}+3H(\rho_{\Lambda}+p_{\Lambda})=-Q.
\end{equation}
In the literature, the usual choices for $Q$ consist of linear terms of either dark matter, dark energy or a linear combination of both. It is possible, however, to consider non-linear forms for $Q$. In \cite{Arevalo2011hh} the cosmological evolution of the universe is analysed, considering a non-linear interaction term of the general form
\begin{equation}\label{e1}
Q=3Hb\rho^{m+s}\rho_{m}^{n}\rho_{\Lambda}^{-s-n},
\end{equation}
where $b$ is a positive coupling constant, $\rho=\rho_m+\rho_{\Lambda}$ and the powers $m$, $n$ and $s$ characterise the interaction. The ansatz (\ref{e1}) contains a variety of interactions considered in the literature, for instance, for $(m,n,s)=(1,1,-1)$ one gets $Q=3Hb\rho_m$ and for $(m,n,s)=(1,0,-1)$ one has $Q=3Hb\rho_\Lambda$.

In this paper we consider the interaction term given by the non-linear form
\begin{equation}\label{eq6}
Q=3Hb\frac{\rho_{\Lambda}^2}{\rho_m+\rho_{\Lambda}},
\end{equation}
which corresponds to the choice $(m,n,s)=(1,0,-2)$ in (\ref{e1}). With this selection we can obtain an analytical expression for the Hubble parameter $H$ which, in addition, is phenomenologically feasible, as we shall show next.

\begin{figure}[t]
\begin{center}
\includegraphics [scale=0.039]{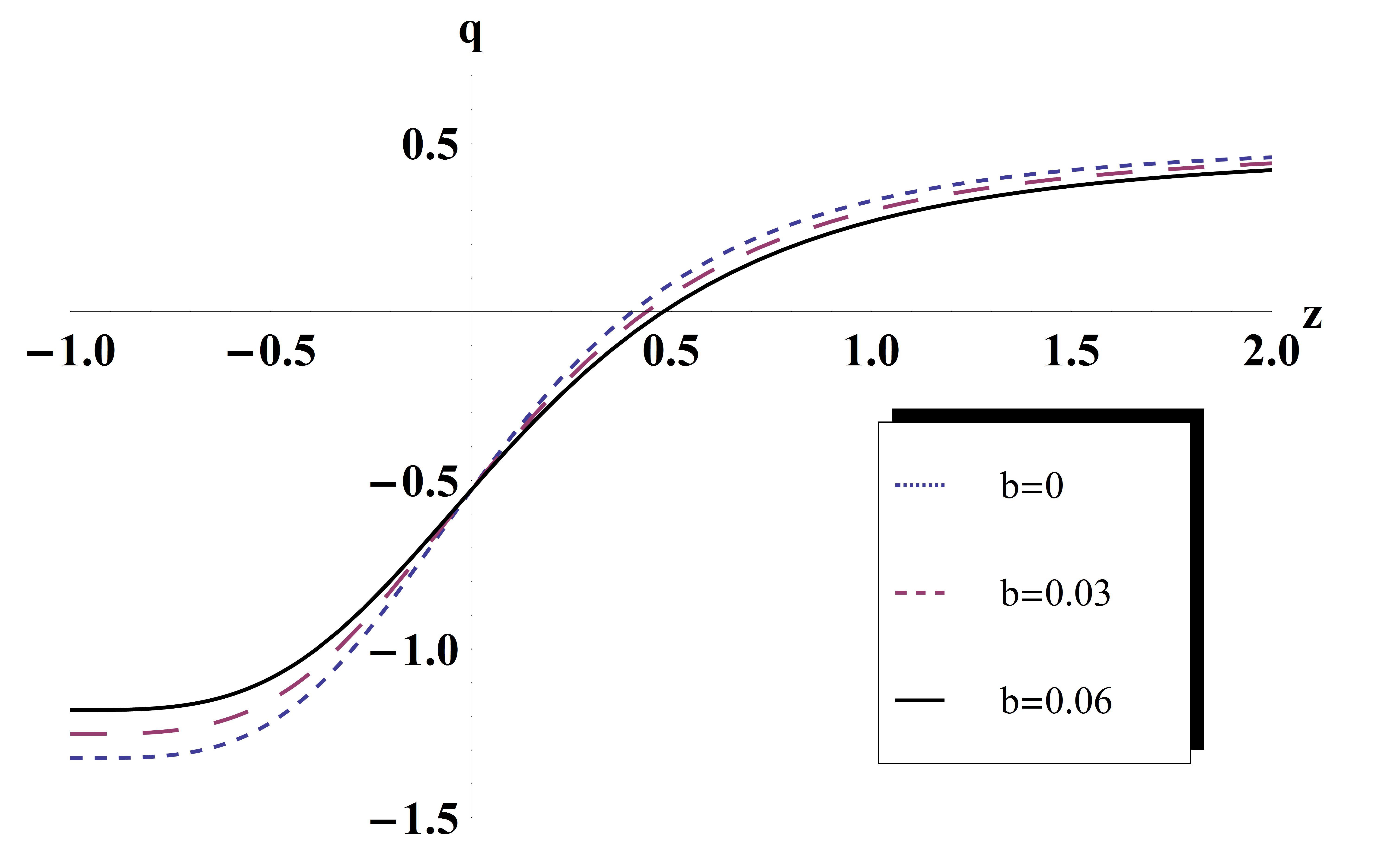}
\end{center}
\caption{Deceleration parameter $q$ as a function of the redshift $z$.}
\label{fig1}
\end{figure}

Replacing (\ref{eq6}) in equation (\ref{eq4}), using $\rho_m$ from equation (\ref{eq1}), an the holographic dark energy density (\ref{eq3}), we obtain the non-linear second order differential equation for $H^2$,
\begin{equation}\label{eq7}
\begin{aligned}
3b\beta^2&\left(\frac{dH^2}{dx}\right)^2 \\ 
         & + 2H^2\left[(6b\alpha\beta+3\beta+2\alpha-2)\frac{dH^2}{dx}+\beta\frac{d^2H^2}{dx^2}\right] \\
         & + 12(b\alpha^2+\alpha-1)(H^2)^2 = 0,
\end{aligned}
\end{equation}
where the change of variable $x = \ln{a}$ has been performed. The solution of (\ref{eq7}) is
\begin{equation}\label{eq8}
\begin{aligned}
& H^2(x)=A\,exp\left[-x\frac{(6b\alpha\beta+2\alpha+3\beta-2)}{\beta(2+3b\beta)}\right] \times \\
& \left[\cos\left(\scriptstyle{\frac{\sqrt{-9(1+4b)\beta^2
-4(\alpha-1)^2+12\beta(2b\alpha+\alpha-1)}} {2\beta}}(x-2\beta B)\right)\right]^{\frac{2}{2+3b\beta}},
\end{aligned}
\end{equation}
with $A$ and $B$ are integration constants, which are recovered from the initial conditions
\begin{equation}\label{eq9}
H^2(0)=H_0^2,
\end{equation}
\begin{equation}\label{eq10}
\left.\frac{dH^2}{dx}\right|_{x=0}=\frac{2(\Omega_{\Lambda 0}-\alpha)H_0^2}{\beta}.
\end{equation}
Here, $H_0$ and $\Omega_{\Lambda 0}$ are the Hubble parameter and the dark energy density at present time ($x=0$, $a=1$), respectively. The initial condition (\ref{eq10}) comes from the dark energy density expression given in (\ref{eq3}). Then, using (\ref{eq9}) and (\ref{eq10}), we get
\begin{equation}\label{eq11}
\begin{aligned}
& A = H_0^2  \big[1 + \\
& \left. \frac{(6b\beta\Omega_{\Lambda 0}+4\Omega_{\Lambda 0}-2\alpha+3\beta-2)^2}{-9(1+4b)\beta^2
-4(\alpha-1)^2+12\beta(2b\alpha+\alpha-1)]}\right]^{\frac{1}{2+3b\beta}}
\end{aligned}
\end{equation}
\begin{equation}\label{eq12}
B=\frac{\tan^{-1} \left[\frac{6b\beta\Omega_{\Lambda 0}+4\Omega_{\Lambda 0}-2\alpha+3\beta-2}{\sqrt{-9(1+4b)\beta^2
-4(\alpha-1)^2+12\beta(2b\alpha+\alpha-1)]}}\right]}
{\sqrt{-9(1+4b)\beta^2
-4(\alpha-1)^2+12\beta(2b\alpha+\alpha-1)}}.
\end{equation}
In order to simplify the next calculations, we rewrite equation (\ref{eq8}) in a compact way, this is,
\begin{equation}\label{eq13}
H^2(x)=A\,e^{-Cx}[\cos D(x-2\beta B)]^{F},
\end{equation}
where $C$, $D$, and $F$ are given by
\begin{equation}\label{eq14}
C=\frac{6b\alpha\beta+2\alpha+3\beta-2}{\beta(2+3b\beta)},
\end{equation}
\begin{equation}\label{eq15}
\begin{aligned}
D &= \frac{1}{2\beta} \times \\ 
  & \sqrt{-9(1+4b)\beta^2 -4(\alpha-1)^2+12\beta(2b\alpha+\alpha-1)},
\end{aligned}
\end{equation}
\begin{equation}\label{eq16F}
F=\frac{2}{2+3b\beta}.
\end{equation}
Clearly, this model has $\alpha$, $\beta$ and $b$ as parameters. Taking into account the known value of the transition redshift $z_t$ in the deceleration parameter $q$, we can perform a fit for $\alpha$ and $\beta$. The deceleration parameter is given by
\begin{equation}\label{eq17}
q=-1-\frac{\dot{H}}{H^2}=-1-\frac{(H^2)'}{2H^2},
\end{equation}
with $(H^2)' = dH^2/dx$. Using eq. (\ref{eq13}), we have
\begin{equation}\label{equ17}
q=\frac{1}{2}[C-2+DF\tan{D(x-2B\beta)}].
\end{equation}
The evolution of $q$ is shown in fig. \ref{fig1}.
The values considered for $b$ were $0.03$ and $0.06$, and also $\alpha = 0.89$, $\beta = 0.34$ and $\Omega_{\Lambda 0} = 0.73$. As can be seen in fig. \ref{fig1}, the selected values for $\alpha$ and $\beta$ produce that $z_t$ takes values of the order of unity, which is consistent with the observed value [\cite{Copeland2006wr}]. These values are the ones we used for the numerical calculations performed from now on.

Considering a barotropic equation of state for the holographic energy density and pressure, the equation of state parameter is $w_{\Lambda}=p_{\Lambda}/\rho_{\Lambda}$, where $\rho_{\Lambda}$ is given by
\begin{equation}\label{equ16}
\rho_{\Lambda}=3\left(\alpha H^2+\frac{1}{2}\beta\frac{dH^{2}}{dx}\right),
\end{equation}
and from eq. (\ref{eq5}), the pressure density $p_{\Lambda}$ is
\begin{equation}\label{eq16}
p_{\Lambda}=-b\frac{\rho^2_{\Lambda}}{\rho_{m}+\rho_{\Lambda}}-\rho_{\Lambda}-\frac{1}{3}\frac{d\rho_{\Lambda}}{dx}.
\end{equation}
Then,

\begin{equation}\label{ecu16}
\begin{aligned}
w_{\Lambda} = &\{\beta(-2D^2F+C(-6+C(2+3b\beta)))   \\
              &-4(-3+C+3bC\beta)\alpha+12b\alpha^2 \\
              &+DF\tan{D(x-2B\beta)} \\
              &\quad \times[-4\alpha+2\beta(-3+C(2+3b\beta)-6b\alpha) \\
              &\quad \quad +D\beta(-2+F(2+3b\beta))\tan{D(x-2B\beta)}]\} \\
              &/\{6(C\beta-2\alpha+DF\beta\tan{D(x-2B\beta)})\}.
\end{aligned}
\end{equation}

\begin{figure}[b]
\begin{center}
\includegraphics [scale=0.039]{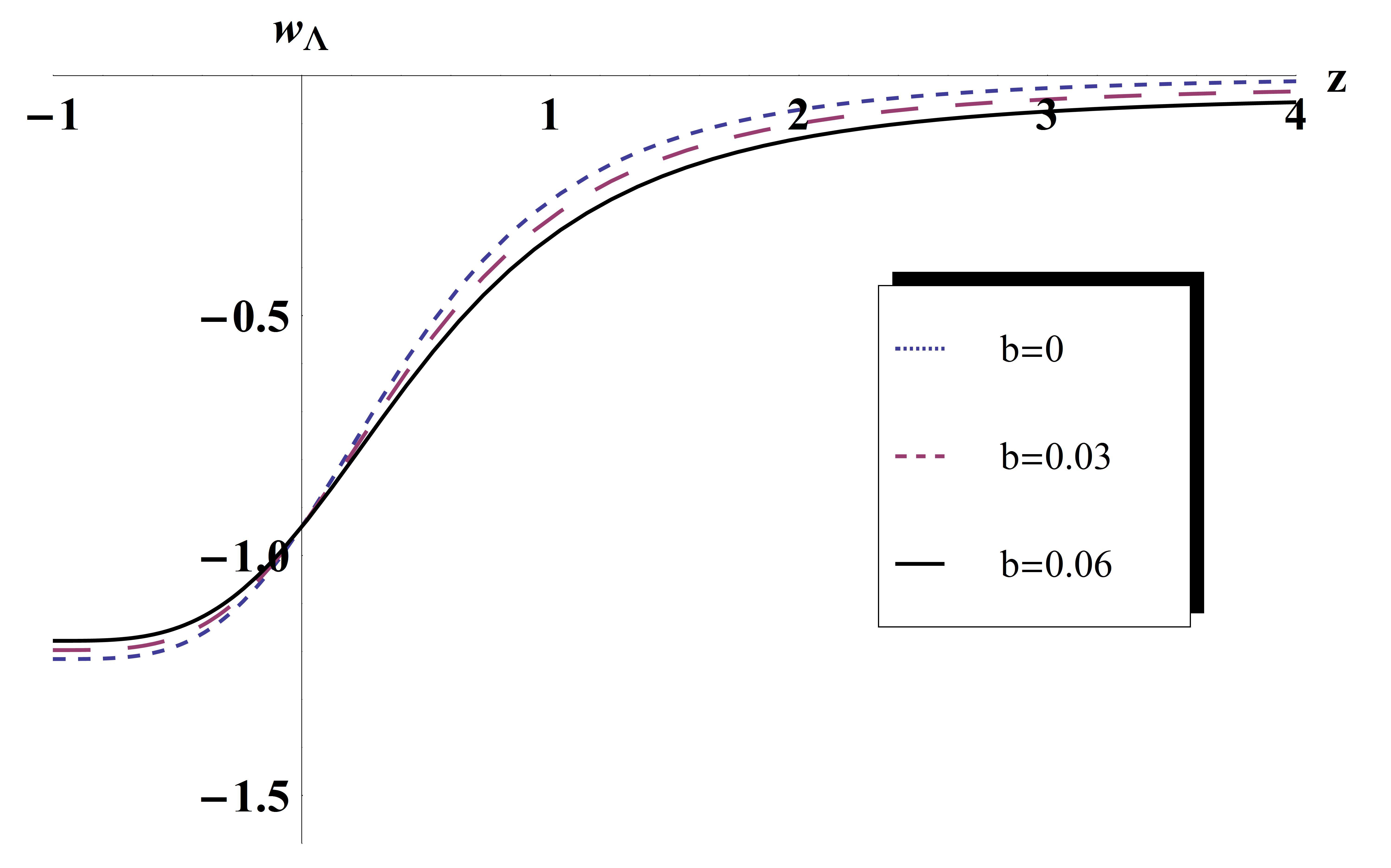}
\end{center}
\caption{Equation of state parameter $w_{\Lambda}$ vs.~redshift $z$.}
\label{fig2}
\end{figure}

Figure \ref{fig2} shows the behaviour of $w_{\Lambda}$ as a function of the redshift $z$.

One can see in figure \ref{fig2} that the equation of state parameter crosses the $-1$ barrier. In this case, the universe evolves in the same way as a scalar fiel known as phantom [\cite{Copeland2006wr}]. This kind of behaviour, where the equation of state takes values grater than $-1$ and also less than $-1$ is typical of a kind of model called quintom [\cite{Feng2004ad,Feng2004ff}]. One can also notice that by increasing the interaction parameter $b$, $w_{\Lambda}$ take more negative values, bellow the $-1$ barrier.

\begin{figure}[t]
\begin{center}
\includegraphics [scale=0.039]{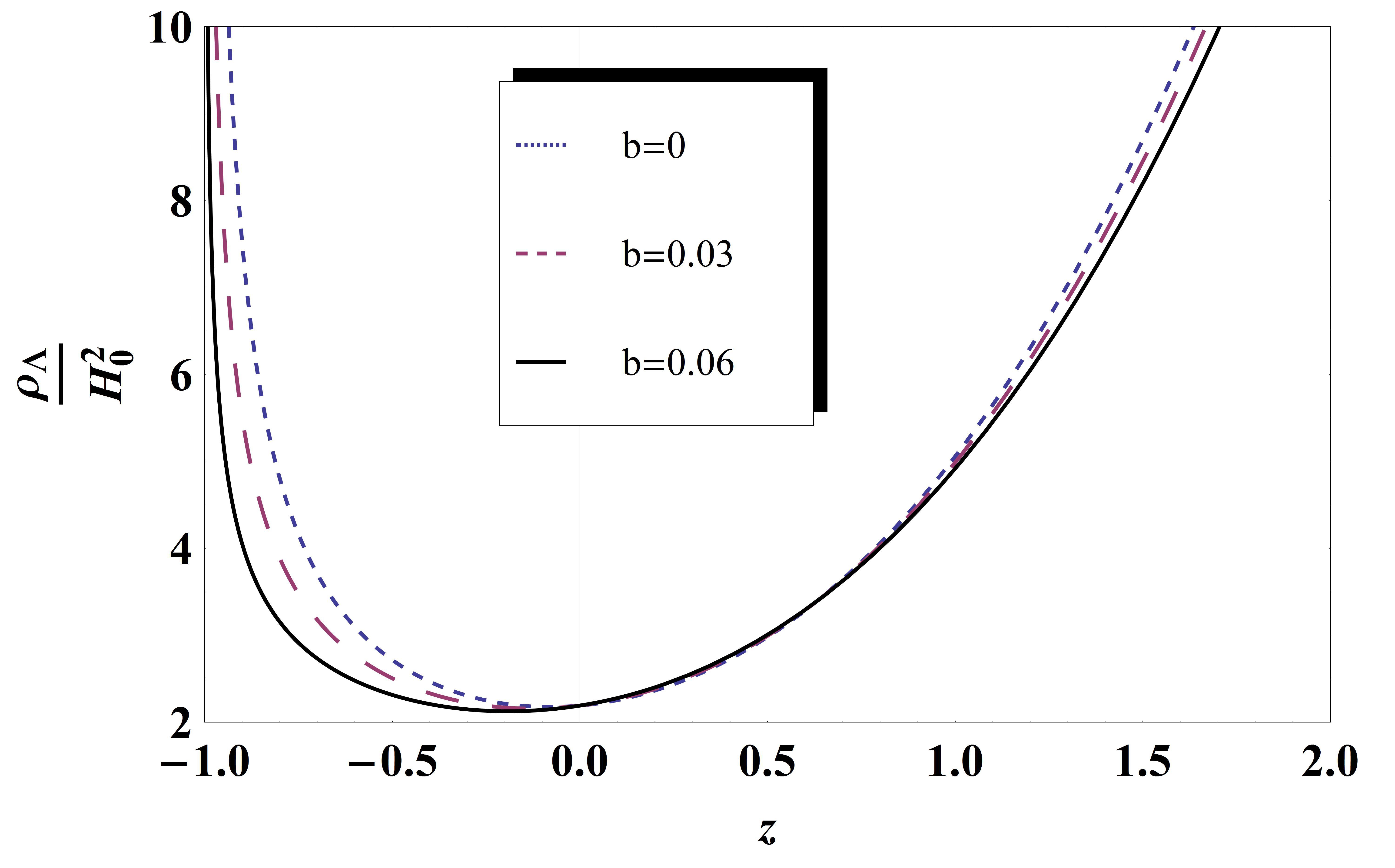}
\end{center}
\caption{Dark energy density $\rho_{\Lambda}$ as a function of the redshift $z$.}
\label{fig3}
\end{figure}

The explicit form for the dark energy density $\rho_{\Lambda}$ is obtained by replacing (\ref{eq13}) on (\ref{equ16}), giving
\begin{equation}\label{equa16}
\begin{aligned}
\rho_{\Lambda} = &-\frac{3}{2}Ae^{-Cx}[\cos{D(x-2B\beta)}]^F \\
                &\times[C\beta-2\alpha+DF\beta\tan{D(x-2B\beta)}].
\end{aligned}
\end{equation}
In fig.~\ref{fig3} the development of $\rho_{\Lambda}$ as a function of the redshift $z$ is shown.
This evolution exhibits a \textquotedblleft Big Rip\textquotedblright--type singularity in the future, which remains for the different values of $b$ we used. This kind of singularities are common in those models whose equation of state crosses the $-1$ barrier (phantom field type behaviour) [\cite{Ma2007pe,Nojiri2009pf}]. For a classification of finite--time singularities depending upon the form of the equation of state of dark energy see Ref.~[\cite{Nojiri2005sx}]. However, we noticed that, for this model, it is possible to avoid this singularity by choosing different values of the parameters $\alpha$ and $\beta$ (for example, $\alpha>1$ and $\beta<1$). For a detailed description of phantom dark energy models without finite-time future singularity, see, for instance, [\cite{Astashenok2012tv}].

\begin{figure}[t]
\begin{center}
\includegraphics [scale=0.039]{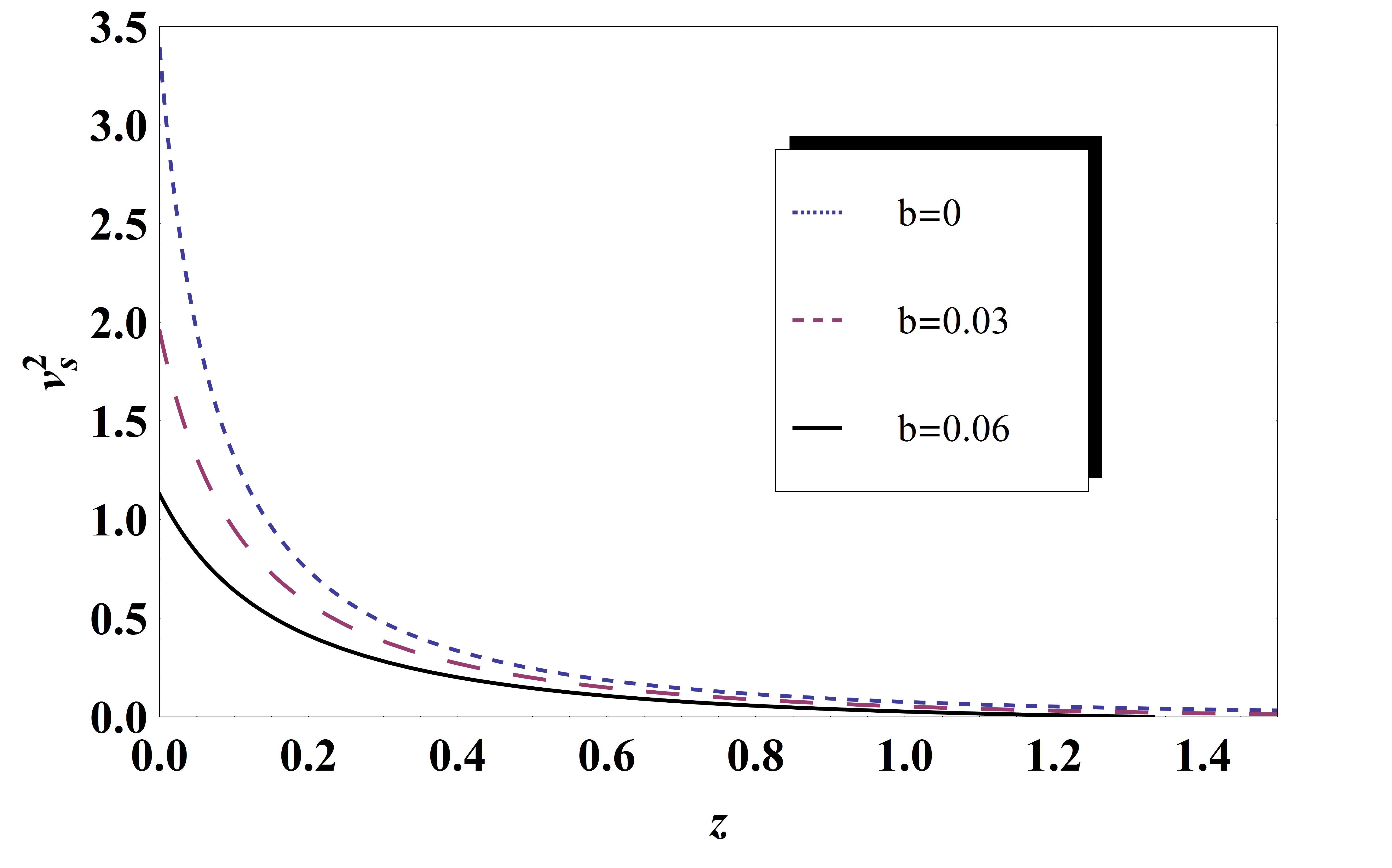}
\end{center}
\caption{Speed of sound squared $v_s^2$ as a function of the redshift $z$.}
\label{fig3b}
\end{figure}

In order to examine the stability of the model, we study the square of the velocity of sound ($v_s^2$) as a function of the redshift. $v_s^2$ is given by [\cite{Peebles2002gy}]
\begin{equation}\label{eqspeed}
v_s^2 = \frac{\dot{p}}{\dot{\rho}} = \frac{p'}{\rho'},
\end{equation}
where the prime means differentiation with respect to $x$. Replacing equations (\ref{equ16}) and (\ref{eq16}) in (\ref{eqspeed}), we generate the plot show in figure \ref{fig3b}. Notice that $v_s^2$ remains positive for all values of $z$, showing that the model presented here is stable under perturbations since the early epoch until present and later time.

\begin{figure}[t]
\begin{center}
\includegraphics [scale=0.039]{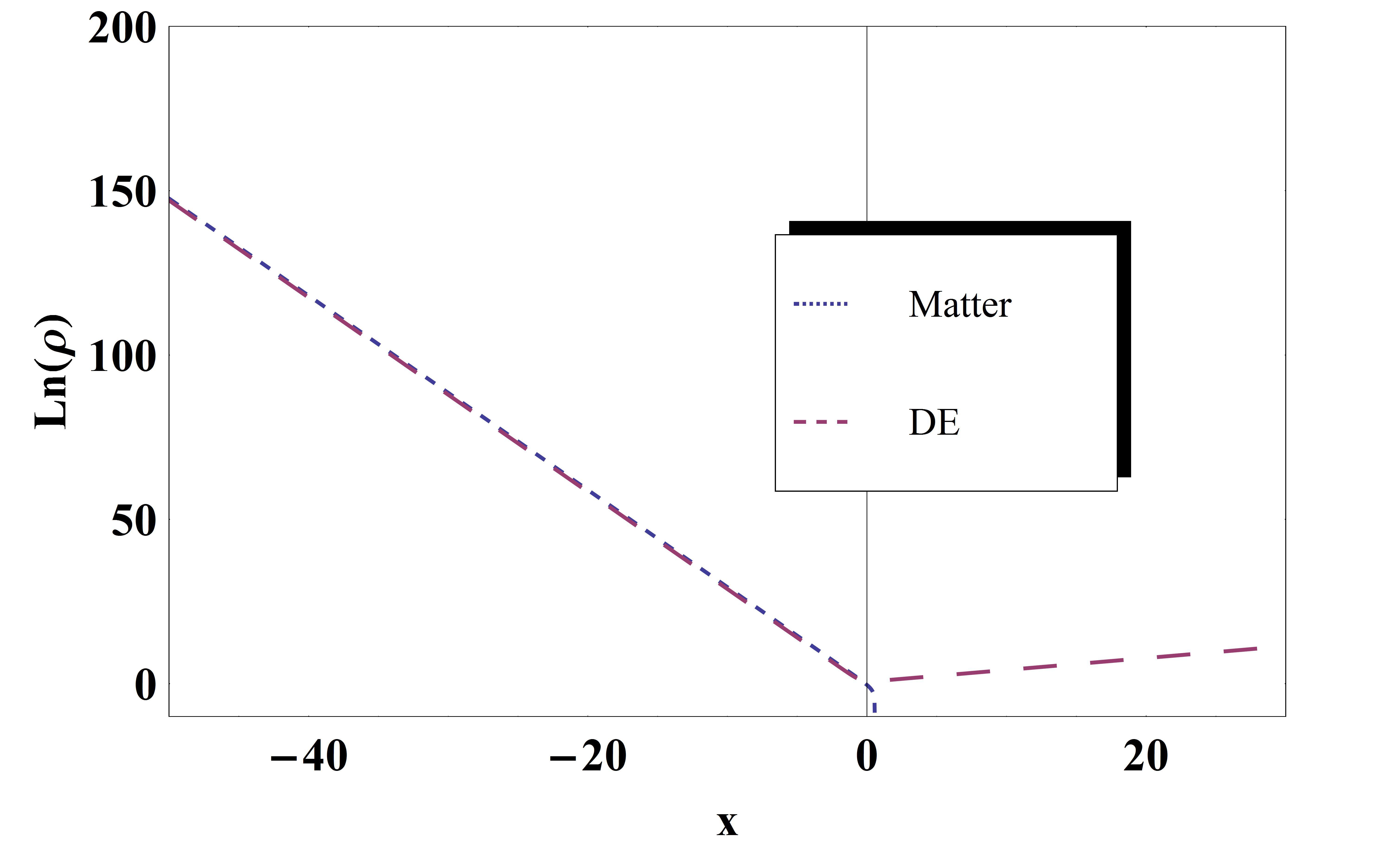}
\end{center}
\caption{Evolution of the Matter and DE densities as a function of $x=\ln{a}$, for $b=0.06$. $x=0$ corresponds to present time.}
\label{fig3c}
\end{figure}

In addition, the model exhibits the possibility to prevent the so--called coincidence problem. As can be seen in Fig.~\ref{fig3c}, the evolution of matter and DE densities remain of the same order in early times; this behaviour changes considerably in the future, when the universe would dilute.

\section[Correspondence of the model with scalar fields]{Correspondence of the model with scalar fields}
In the context of the holographic principle, the authors of [\cite{Bamba2012cp}] have considered scalar field theories and holographic dark energy as models for current acceleration, and demonstrated their equivalence to the corresponding fluid descriptions. Here we consider the correspondence of the holographic model with non-linear interaction with quintessence and tachyon scalar fields.

\subsection[Reconstruction of the model with quintessence]{Reconstruction of the model with quintessence}
In this model a scalar field $\phi$ minimally coupled to the gravitational sector is considered, together with a potential $V(\phi)$ having a shape that will allows, under certain conditions, to get an accelerated expansion regime [\cite{Ratra1987rm,Copeland1997et,Caldwell1997ii,Zlatev1998tr}].

In a FRW flat space--time, the energy and pressure densities for the quintessence scalar field are given by [\cite{Copeland2006wr}]
\begin{equation}\label{eq19}
\rho_q=\frac{1}{2}\dot{\phi}^2+V(\phi),\qquad p_q=\frac{1}{2}\dot{\phi}^2-V(\phi);
\end{equation}
the corresponding equation of state for the scalar field is
\begin{equation}\label{eq20}
w_q=\frac{\dot{\phi}^2-2V(\phi)}{\dot{\phi}^2+2V(\phi)}.
\end{equation}

\begin{figure}[b]
\begin{center}
\includegraphics [scale=0.039]{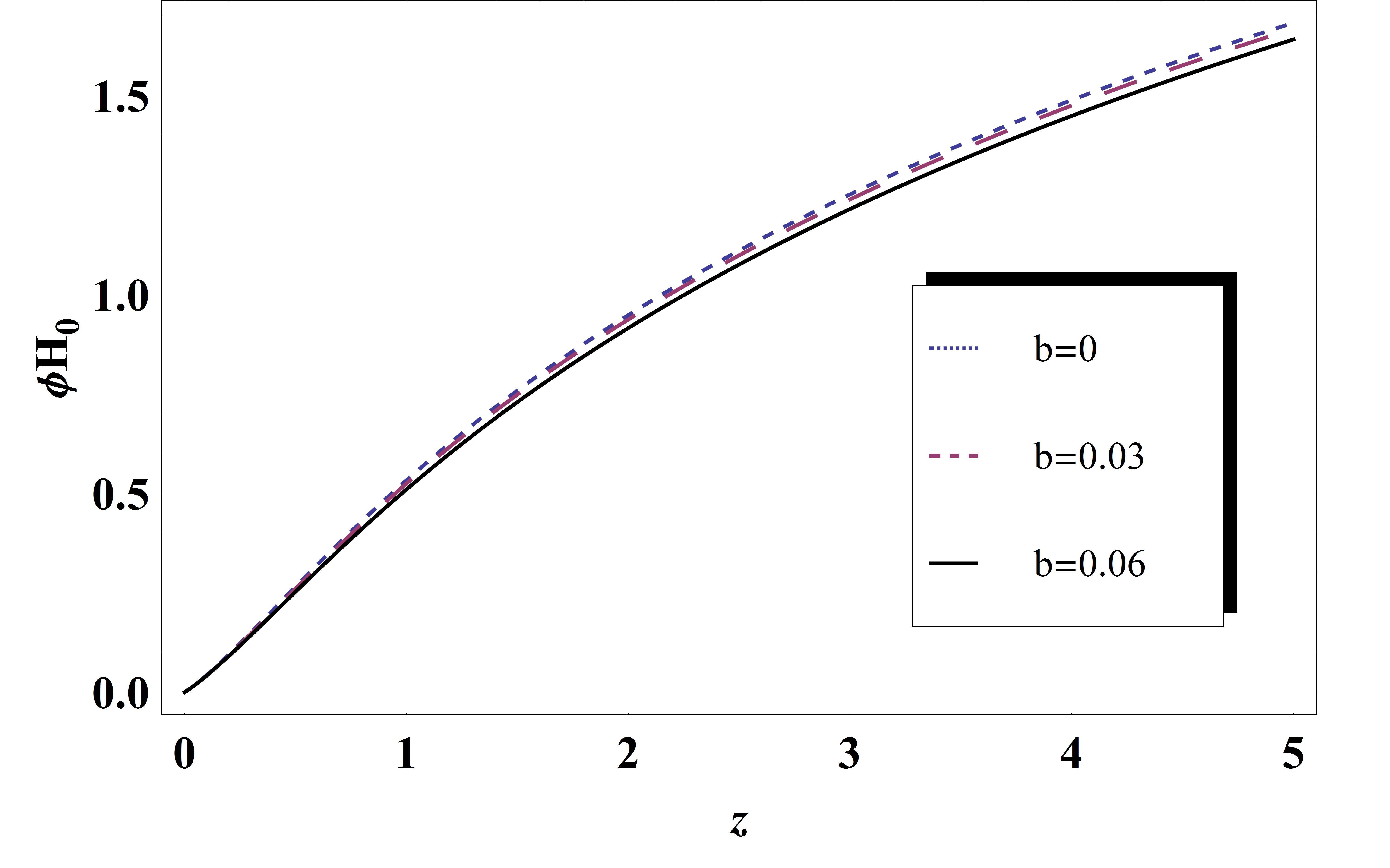}
\end{center}
\caption{Quintessence field evolution as a function of $z$.}
\label{fig4}
\end{figure}

\begin{figure}[t]
\begin{center}
\includegraphics [scale=0.039]{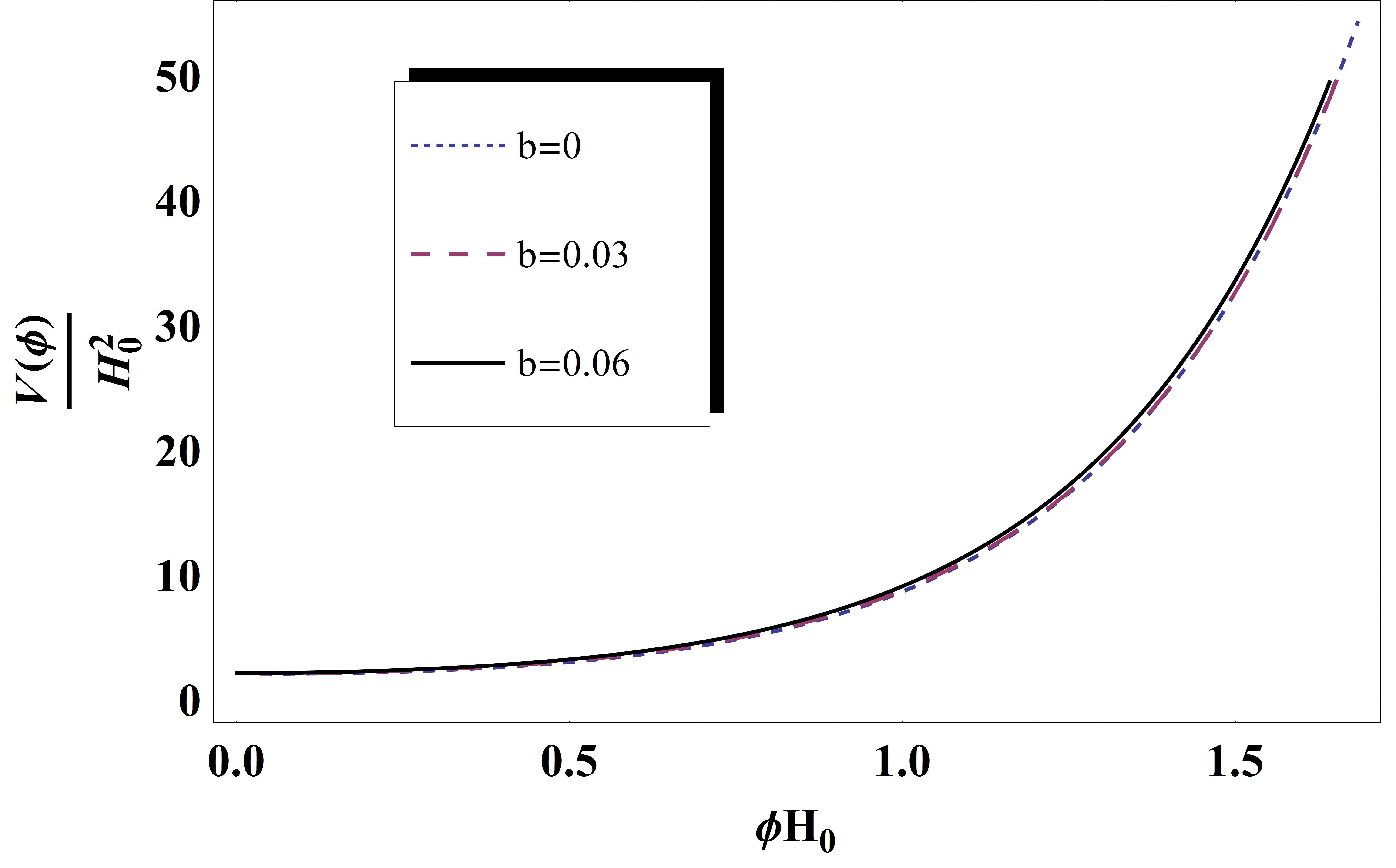}
\end{center}
\caption{Evolution of the quintessence potential as a function of $\phi$.}
\label{fig5}
\end{figure}

To perform the correspondence between the holographic model and the quintessence scalar field, we make $\rho_q=\rho_{\Lambda}$ and $p_q=p_{\Lambda}$. Using eqs.~(\ref{eq13}), (\ref{equ16}), (\ref{eq16}) and (\ref{eq19}), we have
\begin{equation}\label{eq21}
\begin{aligned}
\dot{\phi}^2 &= H^2\left(\frac{d\phi}{dx}\right)^2 = -\frac{1}{4}Ae^{-Cx}[\cos{D(x-2B\beta)}]^F \times\\
\{& -2D^2F\beta + C^2\beta(2+3b\beta) + 12b\alpha^2 - 4C(\alpha + 3b\alpha\beta)\\
  & + DF\tan{D(x-2B\beta)}[2C\beta(2+3b\beta) - 4(\alpha + 3b\alpha\beta)\\
  &\qquad + D\beta(-2 + F(2+3b\beta))\tan{D(x-2B\beta)}]\},
\end{aligned}
\end{equation}

\begin{equation}\label{eq22}
\begin{aligned}
V(\phi) = & \frac{1}{8}Ae^{-Cx}\{\cos{D(x-2B\beta)}\}^F           \\ \times
          & \{-2D^2F\beta+C^2\beta(2+3b\beta)+12\alpha(2+b\alpha) \\
          &-4C(\alpha+3\beta(1+b\alpha)+DF\tan{D(x-2B\beta)}      \\
          &\, \times [6bC\beta^2-4\alpha+4\beta(C-3b\alpha-3) +   \\
          &\quad D\beta(-2+F(2+3b\beta))\tan{D(x-2B\beta)}]\}.
\end{aligned}
\end{equation}
A plot of $\phi$ as a function of $z$ can be obtained by solving numerically the equation (\ref{eq21}), performing the change of variable $x = -\ln{(1+z)}$ and setting $\phi(0)=0$ as the initial condition. The result is shown in figure \ref{fig4}.
The potential $V$ in terms of $\phi$ is displayed in figure \ref{fig5}.

From figure \ref{fig4} is clear that the quintessence field decreases as the universe expands. On the other hand, as we can see from figure \ref{fig5}, the potential is a increasing function of the scalar field $\phi$, showing tendency to be constant (flat) for small values of $\phi$, which is consistent with the accelerated expansion condition, i.~e., $\dot{\phi}\rightarrow 0$ and $w_z\rightarrow -1$. It is also evident from figures \ref{fig4} and \ref{fig5} that the curves are essentially insensitive to variations of the interaction parameter $b$, exhibiting very small variation for late times.

\subsection[Reconstruction of the model with tachyon]{Reconstruction of the model with tachyon}
Another scalar field which has attracted much attention and interest in recent years in the cosmological context is that if tachyon. This kind of fields emerges naturally in the string theory framework [\cite{Padmanabhan2002sh,Bagla2002yn,Abramo2003cp,Aguirregabiria2004xd,Guo2004dta,Copeland2004hq}]. In a FRW universe, the energy density $\rho_t$ and the pressure density $p_t$ for this model are
\begin{equation}\label{eq23}
\rho_t=\frac{V(\phi)}{\sqrt{1-\dot{\phi}^2}},
\end{equation}
\begin{equation}\label{eq24}
p_t=-V(\phi)\sqrt{1-\dot{\phi}^2}.
\end{equation}
The equation of state parameter for the tachyon is
\begin{equation}\label{eq25}
w_t=\frac{p}{\rho}=\dot{\phi}^2-1.
\end{equation}
Then, the correspondence with the holographic model is achieved doing $\rho_t=\rho_{\Lambda}$ and $p_t=p_{\Lambda}$, from which we obtain the expressions for $\dot{\phi}^2$ and $V(\phi)$:
\begin{equation}\label{ecua26}
\begin{aligned}
\dot{\phi}^2 = &[(C-DF)(C+DF)\beta(2+3b\beta) \\
               & - 4C(1+3b\beta)\alpha+12b\alpha^2\\
& + DF(D\beta(-2+F(2+3b\beta))\sec^2{D(x-2B\beta)}\\
& + 2(C\beta(2+3b\beta)\\
& - 2(\alpha+3b\alpha\beta))\tan{D(x-2B\beta)})]\\
& / [6(C\beta(-2\alpha+DF\beta\tan{D(x-2B\beta)})],
\end{aligned}
\end{equation}
\begin{equation}\label{ecua27}
\begin{aligned}
V(\phi) = 
&-\frac{1}{2}i\sqrt{\frac{3}{2}}\,Ae^{-Cx} \{\cos{D(x-2b\beta)}\}^{\frac{2F-3}{2}}\\
&\times \{(C\beta-2\alpha)\cos{D(x-2B\beta)}\\
&\quad + DF\beta\sin{D(x-2B\beta)}\}^{1/2}\\
&\times \{[\beta(-2D^2F+C(-6+C(2+3b\beta)))\\
&\quad - 4(-3+C+3bC\beta)\alpha+12b\alpha^2]\\
&\quad\qquad\times \cos^2{D(x-2B\beta)}\\
&\quad + DF[D\beta(2F+3bF\beta-2) \sin^2{D(x-2B\beta)}\\
&\quad\quad + (-2\alpha+\beta(-3+2C+3bC\beta-6b\alpha))\\
&\quad\quad\quad\times \sin{D(x-2B\beta)}]\}^{1/2}.
\end{aligned}
\end{equation}

\begin{figure}[b]
\begin{center}
\includegraphics [scale=0.039]{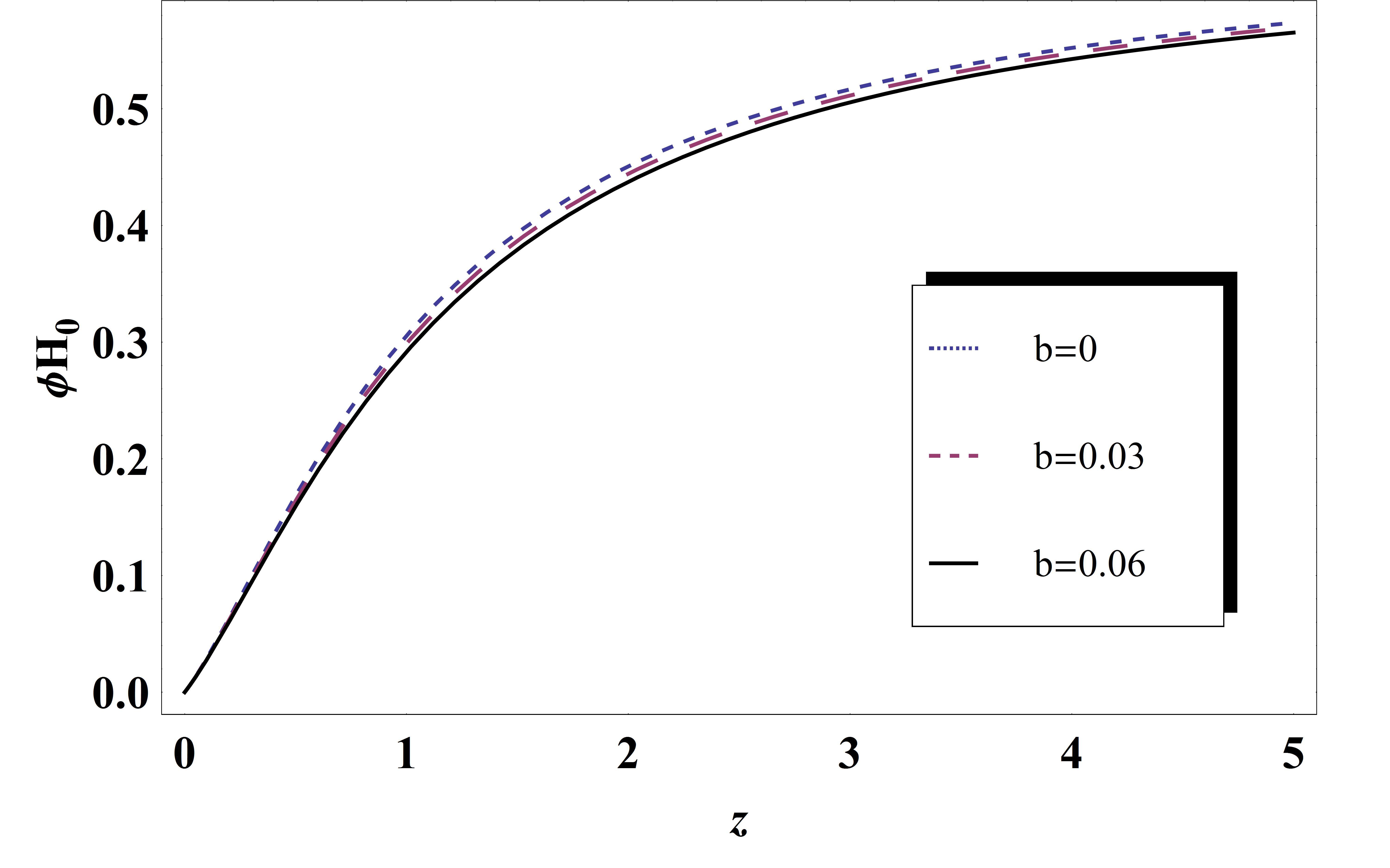}
\end{center}
\caption{Tachyon field evolution as a function of $z$.}
\label{fig6}
\end{figure}

\begin{figure}[t]
\begin{center}
\includegraphics [scale=0.039]{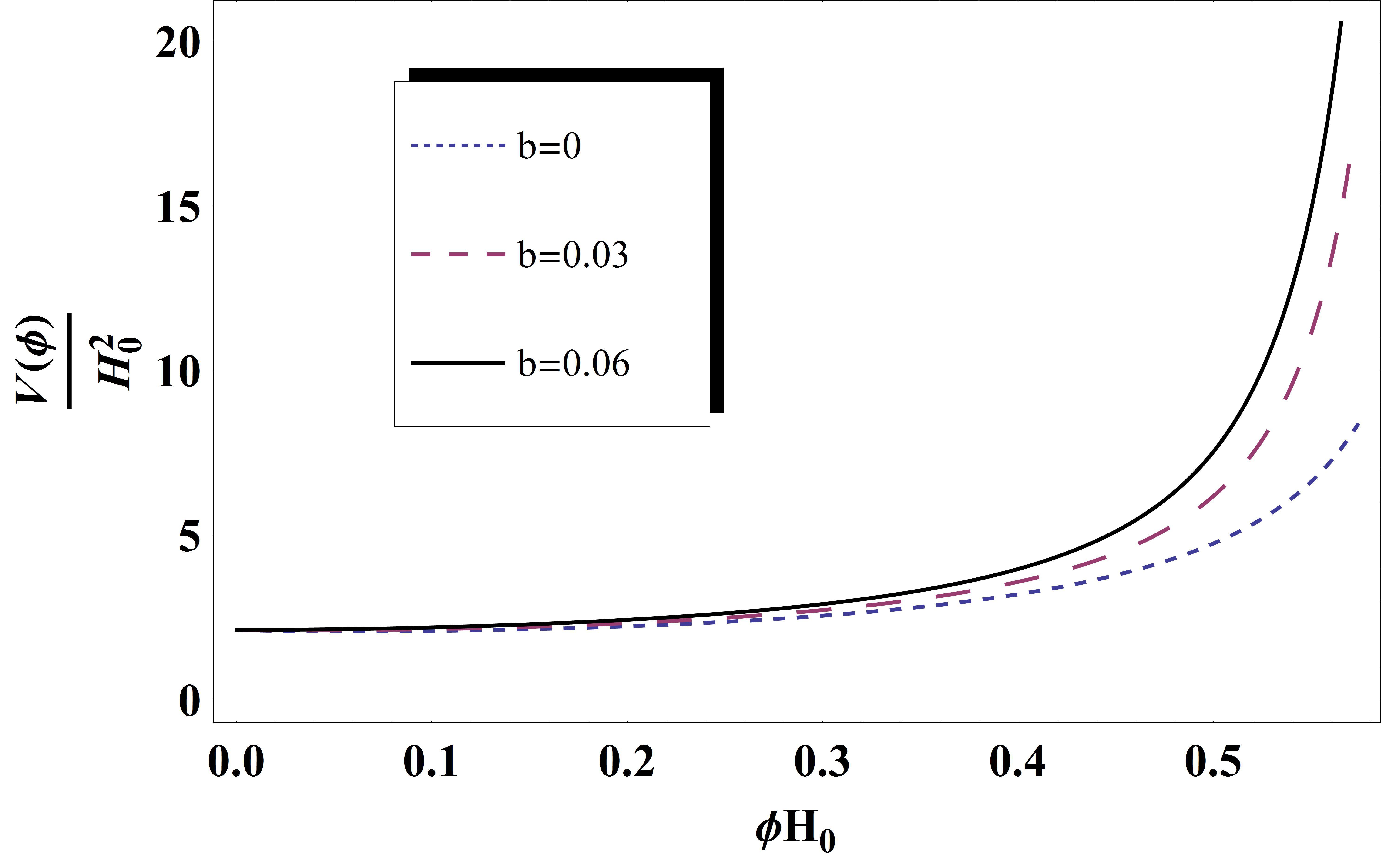}
\end{center}
\caption{Evolution of the tachyon potential as a function of $\phi$.}
\label{fig7}
\end{figure}

It is important to note that the values we use for the parameters $\alpha$ and $\beta$, make $D$ (in equation (\ref{eq15})) to be a complex quantity, so that the potential in (\ref{ecua27}) is real. Figure \ref{fig6} shows the plot of $\phi$ as a function of $z$.
Also, the potential for the tachyon is plotted as a function of the field, and the result is presented in figure \ref{fig7}.
In figure \ref{fig6} one can notice that the tachyon field $\phi$ has a growing behaviour as the values of $z$ increase, but showing a trend to get stabilised for large $z$. Finally, according to figure \ref{fig7}, the potential increases while $\phi$ grows, and for late times ($z$ small), it gets flat and $\dot{\phi}\rightarrow 0$, implying that $w_t\rightarrow -1$, generating an accelerated expansion regime of the universe. This behaviour is similar to the one observed in the quintessence field. Notice also that increasing the interaction parameter $b$, leads to a faster grown of the potential when either $z$ or $\phi$ boosts (see figures \ref{fig6} and \ref{fig7}). 

\section{Conclusions}
In this work we have considered the dark energy holographic model introduced in [\cite{Granda2008dk}], taking into account a dark energy--dark matter non-linear interaction given by equation (\ref{eq6}). In this scenario we found that the dark energy equation of state parameter $w_{\Lambda}$, which shows a quintom--type behaviour, meaning that it takes values above and bellow $-1$ (figure \ref{fig2}) and which corresponds to an accelerated expansion regime ($w_{\Lambda}<-1/3$). The dark energy density $\rho_{\Lambda}$ exhibits a Big-Rip--type singularity in the future ($z\rightarrow -1$), usual for phantom--type models [\cite{Ma2007pe}], which remains for the different values of the interaction parameter $b$ (figure \ref{fig3}).

We also considered the deceleration parameter $q$, and found that, in this context, it presents values for the transition redshift $z_t$ consistent with observations ($z_t\sim 1$) [\cite{Copeland2006wr}] (figure \ref{fig1}) pointing to an appropriate selection of the parameters ($\alpha=0.89$ and $\beta=0.34$).

As shown in fig. \ref{fig3b}, we get positive values for the square of the speed of sound, indicating that the model is stable under perturbations since early times. In addition, the model avoids the coincidence problem as the evolution of the matter and dark energy densities are of the same order for a long period of time (see fig. \ref{fig3c}).

With the correspondence of the model with the quintessence and tachyon scalar fields, we obtained satisfying results. The shape of the potentials are consistent with an accelerated expansion regime for the present universe (figures \ref{fig5} and \ref{fig7}). Similar results are show in the paper by \cite{Sharif2013vga} with the same holographic dark energy model in a Kalusa--Klein universe. 

At last, from our results, we can say that the holographic dark energy model considered here, together with the non--linear interaction term, is a phenomenologically viable model.


\nocite{*}
\bibliographystyle{spr-mp-nameyear-cnd}

\end{document}